\documentclass[journal]{IEEEtran}
\usepackage{cite}
\usepackage{amsmath,amssymb,amsfonts}
\usepackage{algorithmic}
\usepackage{graphicx}
\usepackage{textcomp}
\usepackage{color}
\usepackage{multirow}
\usepackage{booktabs}
\usepackage{makecell}
\usepackage{float}
\usepackage[ruled,linesnumbered]{algorithm2e}
\begin{document}
%
\title{Learning from Pseudo Lesion: A Self-supervised Framework for COVID-19 Diagnosis}
%
%
%


\author{Zhongliang Li, Zhihao Jin, Xuechen Li, Linlin Shen

\thanks{The work is supported by the Natural Science Foundation of China under grants no. 91959108 and U1713214, and the Science and Technology Project of Guangdong Province under grant no. 2018A050501014. Corresponding author: Linlin Shen.}
\thanks{Zhongliang Li, Zhihao Jin, Xuechen Li and Linlin Shen are with the Computer Vision Institute, College of Computer Science and Software Engineering, Shenzhen University, Shenzhen 518060, China, the National Engineering Laboratory for Big Data System Computing Technology, Shenzhen University, Shenzhen 518060, China, and Guangdong Key Laboratory of Intelligent Information Processing, Shenzhen University, Shenzhen 518060, China (email: keloli@163.com; 1900271045,timlee,llshen@szu.edu.cn).}}

\maketitle

\begin{abstract}
The Coronavirus disease 2019 (COVID-19) has rapidly spread all over the world since its first report in December 2019 and thoracic computed tomography (CT) has become one of the main tools for its diagnosis. In recent years, deep learning-based approaches have shown impressive performance in myriad image recognition tasks. However, they usually require a large number of annotated data for training. Inspired by Ground Glass Opacity (GGO), a common finding in COIVD-19 patient’s CT scans, we proposed in this paper a novel self-supervised pretraining method based on pseudo lesions generation and restoration for COVID-19 diagnosis. We used Perlin noise, a gradient noise based mathematical model, to generate lesion-like patterns, which were then randomly pasted to the lung regions of normal CT images to generate pseudo COVID-19 images. The pairs of normal and pseudo COVID-19 images were then used to train an encoder-decoder architecture based U-Net for image restoration, which does not require any labelled data. The pretrained encoder was then fine-tuned using labelled data for COVID-19 diagnosis task. Two public COVID-19 diagnosis datasets made up of CT images were employed for evaluation. Comprehensive experimental results demonstrated that the proposed self-supervised learning approach could extract better feature representation for COVID-19 diagnosis and the accuracy of the proposed method outperformed the supervised model pretrained on large scale images by 6.57\% and 3.03\% on SARS-CoV-2 dataset and Jinan COVID-19 dataset, respectively.
\end{abstract}

\begin{IEEEkeywords}
COVID-19 diagnosis, lesion modeling, self-supervised learning.
\end{IEEEkeywords}

%
\IEEEpeerreviewmaketitle

\section{Introduction}
%
%
%
%

\IEEEPARstart{T}{he} Coronavirus disease 2019 (COVID-19), caused by severe acute respiratory syndrome coronavirus 2 (SARS-CoV-2), has rapidly spread all over the world since its first report in December 2019. According to World Health Organization (WHO),\footnote{https://covid19.who.int/} up to December 2020, more than 70 million cases of COVID-19, including about 1.6 million deaths, have been confirmed all over the world. The reverse transcription-polymerase chain reaction (RT-PCR) assay of sputum or nasopharyngeal swab has been considered as the gold standard for confirmation of COVID-19 cases. However, several studies have reported low sensitivities (i.e., high false negative rates) of RT-PCR for effective early diagnosis and subsequent treatment of presumptive patients\cite{xie2020chest,fang2020sensitivity,ai2020correlation}. In the early stages of the COVID-19 outbreak, the exponentially growing of COVID-19 cases and the serious undersupply of RT-PCR urge us to consider other solutions for COVID-19 diagnosis.\par
Several studies\cite{fang2020sensitivity,bernheim2020chest,lal2020ct,abbasi2020diagnosis} have described typical chest CT imaging findings of COVID-19 as diffuse or focal ground-glass opacities (GGOs). Therefore, thoracic computed tomography (CT), as a non-contact diagnostic method, has become an alternative solution to help diagnose COVID-19 and reduce the risk of infection. However, a CT scan often includes hundreds of slices. The sharp increase in the number of CTs puts radiologists under the burden of onerous diagnosis and leads to fatigue-based diagnostic error. Computer-aided diagnosis (CAD) system is an effective tool to reduce the gap between the tremendous number of CTs and the limited number of radiologist.\par
With the development of deep convolutional neural network (DCNN)\cite{lecun1989backpropagation}, the technology has been widely applied in tremendous challenging tasks in medical image analysis field, e.g. image classification\cite{kermany2018identifying,anthimopoulos2016lung,xu2020novel} and segmentation\cite{li20193d,liu2020ms,zhou2019unet++}. The success of CNNs benefits from the information explosion - the amount of annotated data rapidly increases in the past few years. However, for medical image analysis, the annotated data is difficult to obtain. Take COVID-19 as an example, under the circumstances that medical professionals are highly occupied by taking care of COVID-19 patients, it is difficult to collect and annotate a large number of COVID-19 CT scans for DCNN training. \par
Since deep learning models have a high risk of overfitting when trained on a small-size dataset, transfer learning has been widely used in deep learning-based methods to mitigate the problem of data insufficiency and reduce the risk of overfitting. One commonly used strategy is to learn a powerful deep network for visual feature extraction by pretraining on large datasets in a source task, and then adapt this pretrained network to the target task by fine-tuning on a small-size annotated dataset. This idea has been successfully applied to visual recognition\cite{cao2019feedback} and language comprehension\cite{devlin2018bert}. In the medical image domain, transfer learning has also been widely used in image classification and recognition tasks, such as tumor classification\cite{huynh2016digital,esteva2017dermatologist}, retinal diseases diagnosis\cite{poplin2018prediction}, pneumonia detection\cite{rajpurkar2017chexnet}, lung nodule detection\cite{li2018solitary} and skin lesion classification\cite{oikawa2017pathological,li2018deep}. However, medical images are quite different from nature images. A recent study\cite{raghu2019transfusion} found that the standard large networks pretrained on ImageNet are often over-parameterized and may not be the optimal solution for medical image analysis. As a result, the visual representations learned on ImageNet may not be able to represent CT images, which casts doubts on the transferability from other image sources to COVID-19 CT scans.\par
More recently, as a new paradigm of unsupervised learning, self-supervised learning attracts increasing attentions. The pipeline usually consists of two steps: 1) pretrain a convolutional neural network (CNN) on a manually designed pretext task with a large non-annotated dataset, and 2) fine-tune the pretrained network for the target task with a small annotated dataset. The pretext task enforces CNN to deeply mine useful information from the unlabeled raw data, which can boost the performance of CNN on the subsequent target task with limited training data. The performance is highly related to the design of pretext task. Various pretext tasks have been proposed for self-supervised learning, e.g. colorizing grayscale images\cite{zhang2016colorful}, image inpainting\cite{pathak2016context} and image jigsaw puzzle\cite{noroozi2016unsupervised}, etc. In addition to above mentioned general tasks, a number of downstream task-related pretext tasks have also been proposed and achieved remarkable performance. For example, Vondrick et al.\cite{vondrick2018tracking} leveraged the relationship between contiguous frames to colorize videos and learned a latent tracking feature for visual tracking tasks. Wang et al.\cite{wang2020cycas} forced the model to associate a pedestrian instance to itself after performing a data association between a pair of video frames. The model learned a meaningful representation for the person re-identification problem.\par
Self-supervised learning has also attached increasing interests for medical image analysis and demonstrated impressive performance. Zhou et al.\cite{zhou2020models} and Chen et al.\cite{chen2019self} proposed a set of pretext tasks for anatomical representation learning. However, most of the previous methods were designed for general medical image analysis. Therefore the features cannot well represent specific downstream task like COVID-19 diagnosis.\par
In this work, we propose a novel self-supervised learning strategy, pseudo lesion restoration (PLR), for COVID-19 diagnosis. We used Perlin noise\cite{perlin1985image} to generate the lesion-like patches, which were used to blur the normal CT slices and generate pseudo lesion (PL) images. U-Net\cite{ronneberger2015u} consisting of a pair of encoder and decoder, was trained to restore the PL images to its original version. Then, the encoder was fine-tuned for the COVID-19 diagnosis task. The proposed task was able to learn semantic features closely relevant to COVID-19 diagnosis. When employing hunderds of unlabeled images for network pretraining, our method achieved 96.63\% accuracy on Jinan COVID-19 dataset and 93.9\% accuracy on SARS-CoV-2 dataset, which are 4.71\% and 6.02\% higher than the model pretrained on ImageNet, respectively.\par

\begin{figure}[ht]
\centering
\includegraphics[scale=0.4]{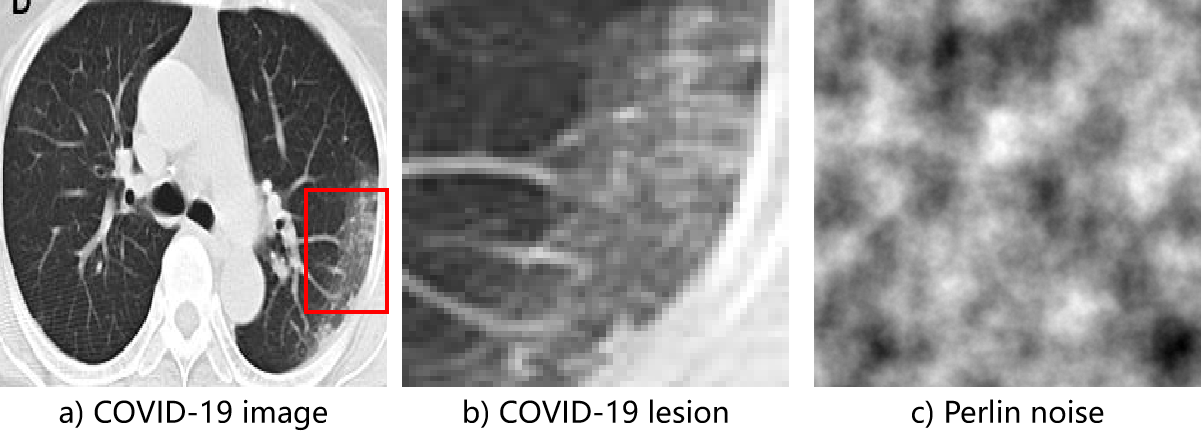}
\caption{Illustration of a) A COVID-19 CT image, b) COVID-19 lesion and c) Perlin noise.}
\label{figl}
\end{figure}

\begin{figure*}[ht]
\centering
\includegraphics[scale=0.43]{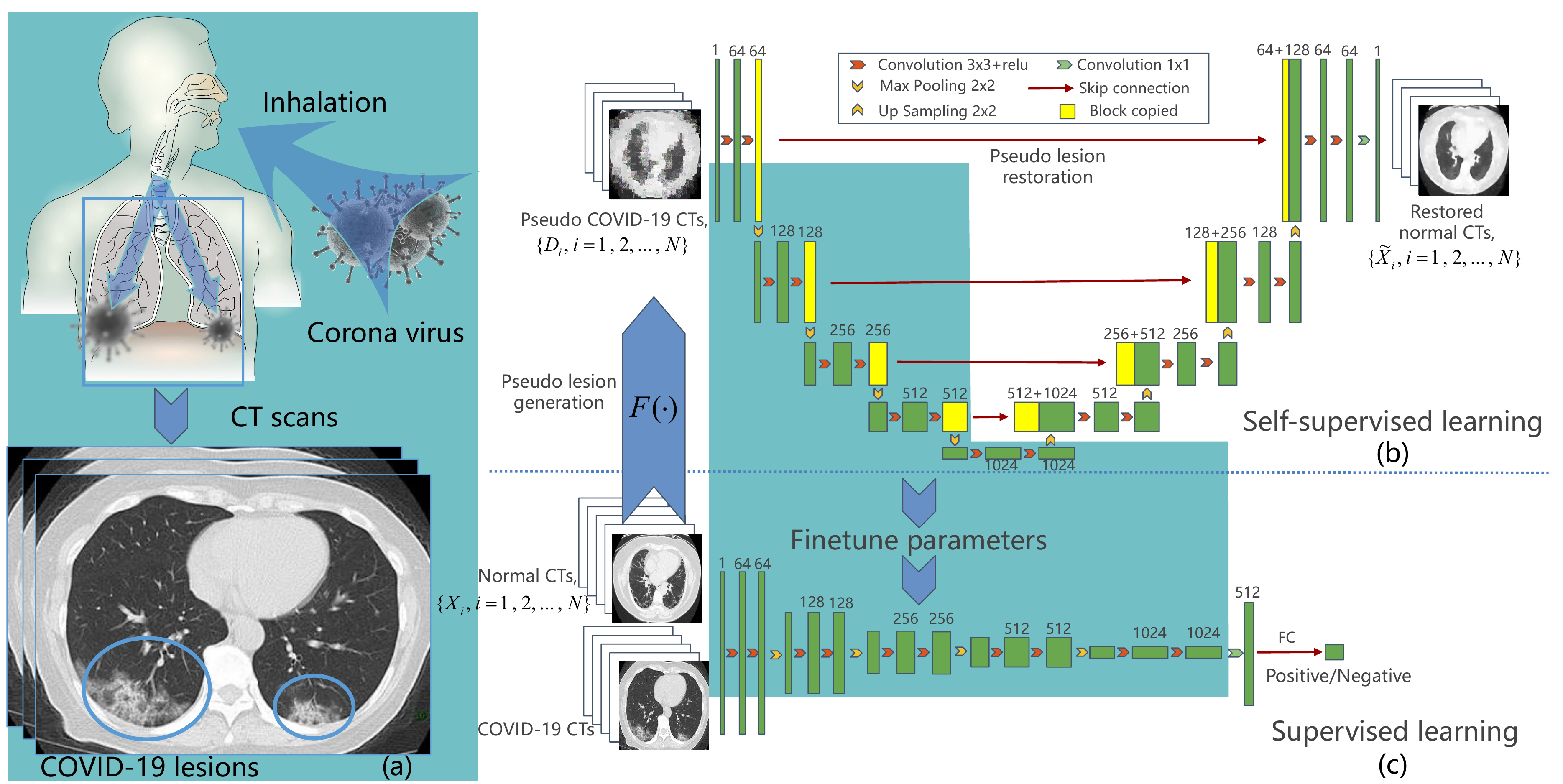}
\caption{The proposed self-supervised pseudo lesion restoration (PLR) framework.}
\label{fig2}
\end{figure*}

The main contributions of this paper are:\par
\begin{itemize}
\item We presented the PLR method, a novel self-supervised pretext task, to learn COVID-19 lesion related representations by restoring normal CT images from pseudo COVID-19 CT images. Our method employed a mathematical model to generate PL, which could be extended for other pulmonary diseases analysis. To the best of our knowledge, this is the first work using self-supervised feature learning for COVID-19 diagnosis.
\item Extensive experiments on two public COVID-19 classification datasets demonstrated the superiority of our method over other supervised and self-supervised methods. Especially, on SARS-CoV-2 dataset, the proposed method achieved 85.1\% accuracy when only 10\% of training set was labeled, which is similar to the performance of random initialization method when the whole training set was labeled.
\item We compared the performance of seven pretraining models for COVID-19 diagnosis using five metrics. The proposed PLR pretraining method achieved the best performance.
\end{itemize}

\section{Related Work}
\subsection{GGO and Perlin noise}
\subsubsection{GGO}
Although the final diagnosis of COVID-19 is based on RT-PCR, the early finding of abnormality in CT are vital for pneumonia detection\cite{kim2020outbreak}. The severe acute respiratory syndrome coronavirus2 (SARS-CoV-2) invades lung tissues through inhaled air\cite{sahin20202019}, which results in the typical imaging finding of ground-glass opacity(GGO). GGO is a descriptive term referring to a region of hazy elevated lung opacity, often fairly diffuse, in which the edges of the pulmonary vessels are difficult to discriminate. GGO is also called “floccus opacity”, as the appearance of GGO is similar to clouds.\par

\subsubsection{Perlin noise}
Simulation of natural scenes e.g. cloud, fire, landscape, etc. is one of the most challenging tasks in computer graphics. In 1985, Ken Perlin present a procedural noise called “Perlin noise”\cite{perlin1985image} for the generation of clouds, waves, incidental motion of animated characters etc. In 2002, Ken Perlin improved Perlin noise using a more efficient interpolation function. In this paper, we use Perlin noise to generate lesion-like patterns as pseudo lesion (PL) patches for model learning. A similar pattern between Perlin noise and COVID-19 lesion can be observed in Fig. 1.\par

\subsection{Deep learning based diagnosis of COVID-19}
Since the outbreak of COVID-19, there have been increasing efforts on developing deep learning based methods\cite{harmon2020artificial,wang2020covid} for medical image based COVID-19 screening. More Recently, a great number of works have proved the potential capacity of deep models for COVID-19 diagnosis and automated segmentation of infection\cite{fan2020inf}. Sun et al.\cite{sun2020adaptive} propose an adaptive feature selection-guided deep forest (AFS-DF) for COVID-19 classification based on chest CT images. Wang et al.\cite{wang2020covid} proposed COVID-Net for the detection of COVID-19 cases from chest radiography images. Wang et al.\cite{wang2020weakly} introduced a weakly-supervised deep learning framework trained on 3D CT volumes for COVID-19 infection localization. Ma et al.\cite{ma2020towards} created a COVID-19 3D CT dataset with 20 cases that contains above 1800 annotated slices and a pretrained segmentation model. Although Shan et al.\cite{shan2020lung} proposed a human-in-the-loop workflow to improve the efficiency of data labeling, the performance of COVID-19 CAD network is still confined by the limited amount of annotated data.\par

\subsection{Self-supervised learning}
To deal with the problem of limited number of annotated data, researchers attempted to exploit useful information from the unlabeled data with unsupervised approaches\cite{spitzer2018improving,zhang2017self}. Self-supervised learning (SSL) aims to learn meaningful representations of input data without manual annotations. It creates auxiliary tasks and forces deep networks to learn highly-effective latent features. Various strategies have been proposed to construct pretext tasks\cite{vondrick2018tracking,wang2020cycas}. For natural images, Doersch et al.\cite{doersch2015unsupervised} proposed a pretext task to predict the relative positions of two patches from the same image. Komodakis et al.\cite{komodakis2018unsupervised} proposed a simple but effective task, to predict the 2D rotation for unsupervised representation learning.\par
For the applications to medical data, Zhang et al.\cite{zhang2017self} defined a pretext task that sorted the 2D axial slices extracted from the conventional 3D CT and MR volumes to pretrain neural networks for the fine-grained body part recognition. Zhu et al.\cite{zhu2020rubik} proposed a self-supervised feature learning framework to learn translation and rotation invariant features from a series of operations e.g. cube ordering, cube rotating and cube masking for 3D medical analysis.\par
Apart from the aforementioned pretext tasks\cite{zhang2017self,zhu2020rubik}, several recent studies\cite{zhou2020models,chen2019self} pretrained neural networks by restoring the content of medical images with a encoder-decoder architecture. For examples, Zhou et al.\cite{zhou2020models} trained a genesis model to learn robust representation by restoring 3D image patches from a set of transformation, e.g. local pixel shuffling, non-linear transformation, outer-cutout, and inner-cutout, etc. Chen et al.\cite{chen2019self} proposed a self-supervised learning strategy based on context restoration, which enables CNNs to learn useful image semantics and benefits subsequent tasks. However, as a vital finding in COVID-19 CTs, GGO has not been considered in pretext task design to learn semantic features for COVID-19 diagnosis.

\section{Method}
\subsection{Overview of the PLR framework}
To address the large annotation cost of COVID-19 CT scans, we proposed a novel self-supervised learning approach, pseudo lesion restoration (PLR), to learn more robust features for COVID-10 diagnosis. The pipeline of PLR approach is illustrated in Fig. 2.
As shown in Fig. 2(a), COVID-19 lesions presents the pattern of floccus  opacities. Thus, we employed Perlin noise to generate pseudo lesion-like patches and designed a restoration task to remove these lesion-like noises. The image restoration task was implemented with an encode-decoder structure trained by paired input/output images (Fig. 2(b)). After PLR pretraining, the encoder alone is fine-tuned for target classification task with annotated CT images (Fig. 2(c)).
\subsection{Pseudo lesion and COVID-19 image generation}
In this work, 500 noise images with size 512$\times$512 was generated based on Perlin noise, which were used to create lesion-like patches. Then the lesion-like patches were pasted to normal CT images to generate PL images.\par
Perlin noise is a gradient noise that is built from a set of pseudo-random gradient vectors of unit length evenly distributed in N-dimensional space. Unlike regular noise, Perlin noise use some coherent structure to generate ‘realistic’ structures.\par

\begin{figure}[ht]
\centering
\includegraphics[scale=0.3]{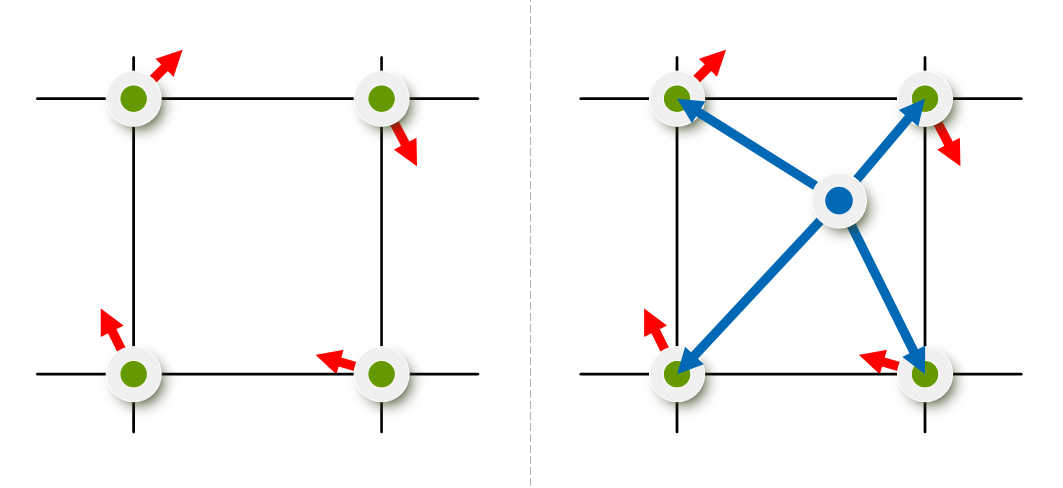}
\caption{Left: grid points with assigned gradient vectors (red), right: vectors to grid vertices (blue).}
\label{fig3}
\end{figure}

\begin{figure}[ht]
\centering
\includegraphics[scale=0.5]{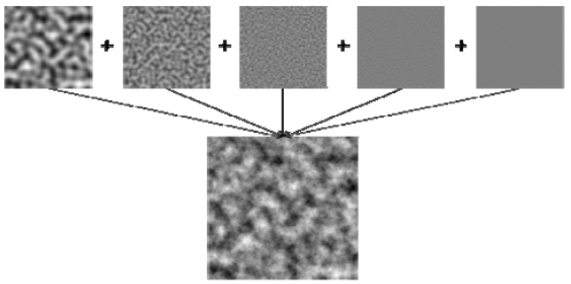}
\caption{An example noise image generated from Perlin noises with different noise frequencies and amplitudes.}
\label{fig4}
\end{figure}

\begin{figure}[ht]
\centering
\includegraphics[scale=0.29]{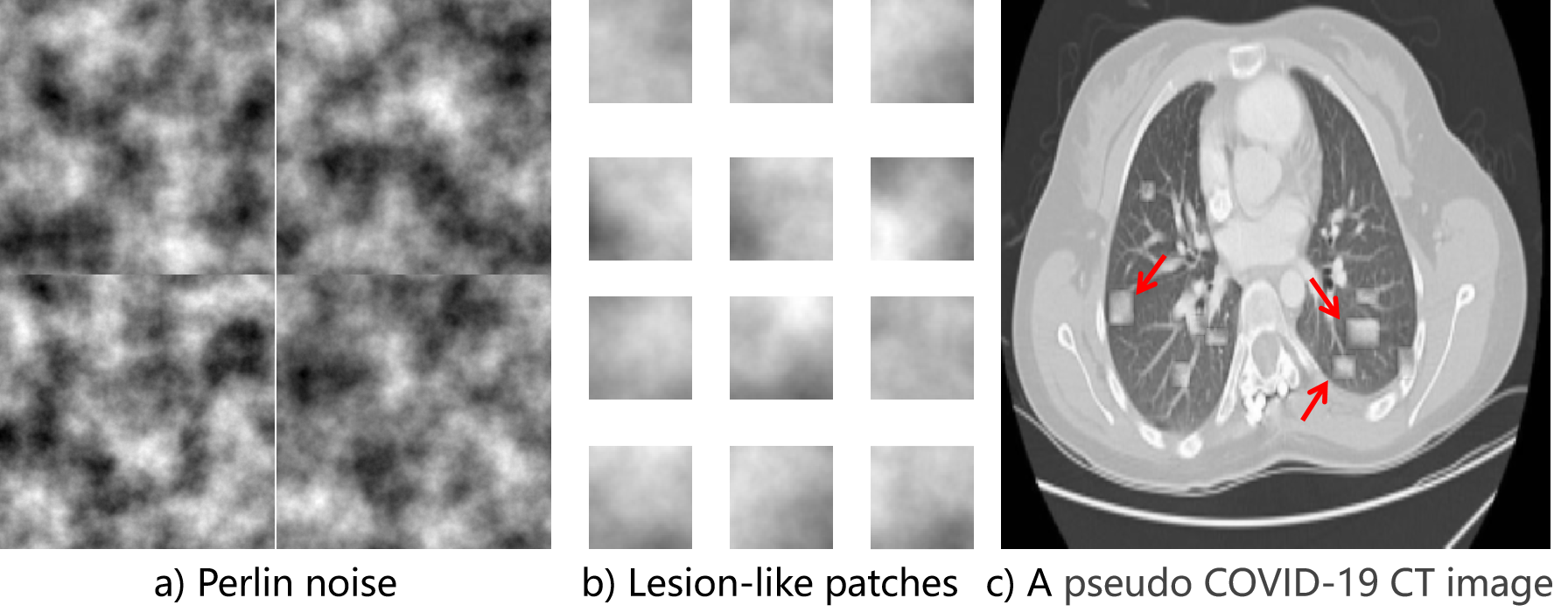}
\caption{Illustration of a) Perlin noise images, b) lesion-like patches and c) a pseudo COVID-19 CT image.}
\label{fig5}
\end{figure}

In the two-dimensional space, Perlin noise at point $(x,y)$ is determined by computing a pseudo-random gradient at each of the four nearest vertices on the integer grid and then spline interpolated. Let $(i,j)$ denotes the four points on this grid, where $i$ is the set of lower and upper bounding integers on $i={\lfloor x \rfloor,\lfloor x \rfloor+1}$, and similarly $j={\lfloor y \rfloor,\lfloor y \rfloor+1}$. The four gradients are given by $g_{i,j}=G[P[P[i]+j]]$ where precomputed arrays $P$ and $G$ contain a pseudo-random unit-length gradient vectors, respectively. The successive application of $P$ hashes each grid point to de-correlate the indices into $G$. The four linear functions $g_{i,j}·(x-i,y-j)$  (Fig. 3(b)) are then trilinearly interpolated by $s(x-\lfloor x \rfloor), s(y-\lfloor y \rfloor)$, where $s(t)=6t^5-15t^4+10t^3$.\par

Perlin noise combines multiple functions called ‘octaves’ to produce natural looking surfaces. Each octave adds a layer of detail to the surface. For example, octave 1 could be “mountains”, octave 2 could be “boulders”, and octave 3 could be the “rocks”. Fig. 4 shows a noise image consisting of basic Perlin noise with different frequency and amplitudes.\par
As the grey level intensities of GGO are usually high, the set of Perlin noise images are further divided into small patches with random sizes (width and height:15±10 pixels), and 5000 lesion-like patches with average grey intensity larger than 180 are selected as the candidates for PL image generation. For each normal CT image $X_i$, a random number of lesion-like patches were pasted to the random locations within lung region to generate a PL image $D_i$.  Fig. 5 shows the example of noise images, lesion-like patches and a PL image generated by pasting lesion-like patches to a normal CT scan.\par

We also employed Gaussian blur and local pixel shuffling for pseudo lesion generation. To be specific, we used $5\times5$ Gaussian kernels with $\sigma=0$ to blur each normal CT image $X_{i}$ for pseudo COVID-19 image generation. For local shuffling, we divided $X_{i}$ to 100 image patches, then shuffle the pixels inside each patch sequentially. The Perlin noise based, Gaussian blurred or local pixel shuffled normal CT images $\lbrace X_i,i=1,2,…,N \rbrace$ are considered as pseudo COVID-19 images $\lbrace D_i,i=1,2,…,N \rbrace$ for PLR-P, PLR-G and PLR-L model pretraining, respectively.\par

\subsection{Self-supervised Pseudo lesion restoration}
Given a set of $N$ PL images $\lbrace D_i,i=1,2,…,N \rbrace$ generated from normal CT images $\lbrace X_i,i=1,2,…,N \rbrace$, our pretext task is to restore $X_i$ from $D_i$. In this work, we adopt U-Net\cite{ronneberger2015u} as the encoder-decoder structure for image restoration and apply the encoder for the following COVID-19 diagnosis task.
As shown in Fig. 2(b), our U-Net consists of stacks of convolutional units and downsampling units for feature extraction and reconstruction. The restoration part used a number of convolutional layers and upsampling layers to generate the restored normal CT images $\lbrace \widetilde{X}_i,i=1,2,…,N \rbrace$. Skip connections were used to propagate multi-scale features to multi-resolution layers. Mean Squared Error (MSE) loss was employed for the training of PLR network.
While previous self-supervised approaches design general proxy tasks like local pixel shuffling and context restoration for semantic image features learning, our task is specific to COVID-19 lesions. The network is expected to learn deep features related to GGO and can thus benefit the following COVID-19 classification tasks. 

\subsection{Encoder finetuning for COVID-19 classification}
Once the encoder-decoder network has been trained using the PLR task, the network can learn representation of lesion-like patterns, which can be transferred to the COVID-19 diagnosis task. Our classification network was designed by simply concatenate the encoder of the pretrained U-Net with a FC layer of 512 neurons (Fig. 2(c)). The limited number of annotated CT images were used to finetune the classification model.

\subsection{Implementation}
The PLR framework is built on Keras deep learning framework with an NVIDIA Tesla P100 32G GPU on Ubuntu 16.04.4 LTS system. Hyper-parameters are finetuned on the validation set. For all pretrained models and finetuned models, the raw image intensities were normalized to the range of $[0,1]$ before training. For image restoration tasks, tampered normal images were resized to 512$\times$512 and PLR network were trained with a mini-batch size of 4, using stochastic gradient descent (SGD) method with an initial learning rate of 1e-3. We use ReduceLROnPlateau to schedule learning rate, i.e. if there is no accuracy improvement in the validation set for a certain number of epochs, the learning rate is reduced. For image classification tasks, models are trained with a min-batch size of 16, Adamdelta optimizer\cite{zeiler2012adadeltaa} with an initial learning rate of 0.1 is used for optimization. All training images are augmented using random zooming and sheering with a random factor chosen uniformly from $[0.8, 1.2]$. The images were resized to 224$\times$224 for SARS-CoV-2 dataset. Specially, to fairly compare the results achieved by the previous work, the input images are resized to 512$\times$512 for Jinan COVID-19 dataset.

\begin{figure}[ht]
\centering
\includegraphics[scale=0.35]{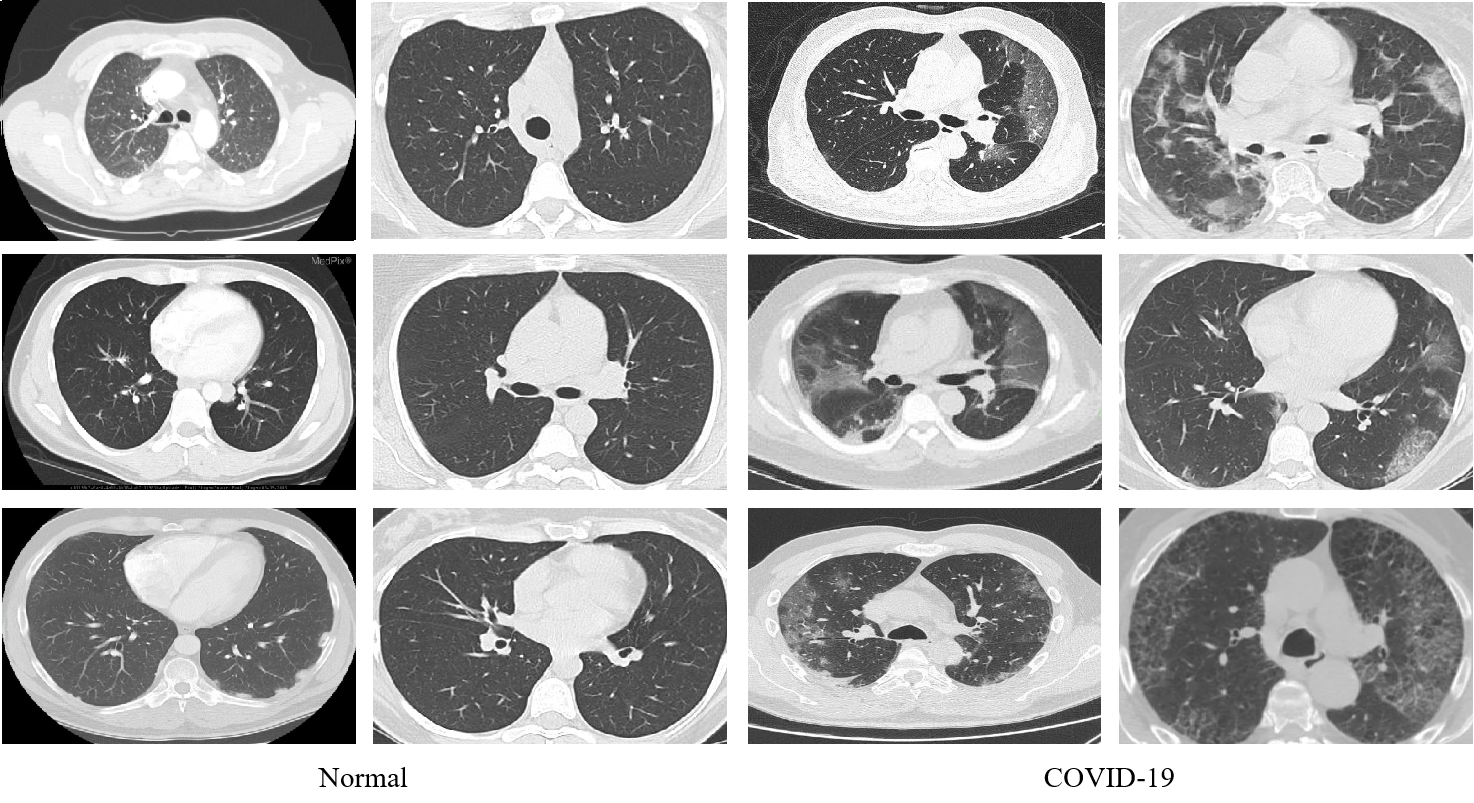}
\caption{The CT images of normal and COVID-19 patients from two different public datasets, showing data heterogeneity on the appearance and contrast. The first two columns show normal CT images, while the remaining two columns show COVID-19 CTs. From left to right, column (1), (3) illustrate CT images in Jinan COVID-19 dataset and column (2), (4) illustrate CT images in SARS-CoV-2 dataset.}
\label{fig6}
\end{figure}

\begin{table}[]
\centering
\caption{The details of Training, Validation and Test Set for two datasets.}
\label{tab:table1}
\begin{tabular}{lllll}
\hline
Dataset                         & Class    & Training & Validation & Test \\ \hline
\multirow{2}{*}{Jinan COVID-19} & COVID-19 & 309      & 46         & 50   \\
                                & Normal   & 303      & 45         & 49   \\ \hline
\multirow{2}{*}{SARS-CoV-2}     & COVID-19 & 1295     & 253        & 619  \\
                                & Normal   & 464      & 76         & 217  \\ \hline
\end{tabular}
\end{table}

\section{Experimental Results}
In this section, we firstly introduce two public COVID-19 classification datasets and the evaluation metrics. Then, we describe the structure of the networks and the setting for experiments in detail. Finally, we illustrate the results of image restoration, and more importantly, comparisons with five different pretraining strategies.\par
\subsection{Datasets and Evaluation Metrics}
Two public COVID-19 CT datasets, i.e. Jinan COVID-19 dataset\cite{hu2020automated} and SARS-CoV-2 dataset\cite{soares2020sarscov2}, were employed to evaluate the proposed PLR method and other pretraining strategies. Jinan COVID-19 dataset \footnote{https://github.com/KevinHuRunWen/COVID-19/} contains 802 chest CT images, which are divided into two classes, i.e. 405 COVID-19 and 397 Normal. The dataset was created by collecting images from COVID-19 related works published in medRxiv, bioRxiv, NEJM, JAMA, Lancet, and other journals. SARS-CoV-2 dataset was collected from Sao Paulo, which consists of 2925 CT images from 130 patients, in which 2167 are positive with COVID-19 and 758 are normal. It should be stressed here that we the employed the second version of SARS-CoV-2 dataset, because label leaking exist in its first version. The details of training, validation and testing dataset for the two COVID-19 dataset were summarized in Table \uppercase\expandafter{\romannumeral1} and samples in the datasets were given in Fig. 6.\par 
We employed five widely used metrics, i.e. accuracy, precision, recall, F1 score and AUC to quantitatively evaluate the classification performance of different models. We define “COVID-19” as positive, “Normal” as negative, TP, TN, FP, and FN as true positives, true negative, false positives, and false negatives, respectively. AUC represent the area under ROC curve. Following is the formula of metrics for evaluating model performance:\\

\begin{equation}
    Accuracy = \frac{TP+TN}{TP+TN+FP+FN}
\end{equation}
    
\begin{equation}
    Precision = \frac{TP}{TP+FP}
\end{equation}

\begin{equation}
    Recall = \frac{TP}{TP+FN}
\end{equation}

\begin{equation}
    F1~score = \frac{2\times Precision \times Recall}{Precision+Recall}
\end{equation}

\subsection{Experimental settings}
\subsubsection{Backbone network}
The U-Net is adopted for image restoration (Fig. 2(b)) in this work. The encoder of U-Net, denoted as E-U-Net, consists of 10 convolutional layers with $3\times3$ kernels. A Global Average Pooling (GAP) layer was employed for converging convolutional features learned in E-U-Net, and followed by two fully connected layers. Each convolution operation was activated by ReLU function. Mean Squared Error (MSE) and binary cross-entropy losses were employed to supervise the image restoration and classification tasks, respectively.\par

\subsubsection{Baseline overview}
For a through comparison, we compared the proposed method with five network initialization methods, including (1) a random initialization (RI) method\cite{he2015delving}, (2) a supervised pretraining with ImageNet\cite{krizhevsky2017imagenet} and (3) a supervised pretraining with PneuXray dataset\cite{li2019pnet}. While most of the previously proposed methods adopt rotation or relative patch position prediction as pretext task to extract contextual information, they might not be the appropriate solution for COVID-19 lesion representation. We employed “pseudo lesions” to establish relationships between normal and COVID-19 CT images, and set restoration as the pretext tasks. In addition to supervised approaches, two self-supervised methods i.e. (4) restoring normal CT images from local pixel shuffled (LPS)\cite{zhou2020models} CT images and (5) restoring CT images from Gaussian blurred (GB) CT images are also involved for comparison. Especially, to promote the self-supervised deep models to learn more robust features, 1000 LPS and 500 GB operations were employed in method (4) and (5), respectively. In table \uppercase\expandafter{\romannumeral2}, we summarized the network initialization methods in details.\par

\begin{table}[]
\centering
\caption{The details of network initialization methods.}
\label{tab:table2}
\begin{tabular}{l|l}
\hline
\begin{tabular}[c]{@{}l@{}} Methods\end{tabular} & Description                                                                                                                                                                                  \\ \hline
RI                                                                         & \begin{tabular}[c]{@{}l@{}}A random initialization (RI) method, called \\ MSRA,was suggested in He et al.\cite{he2015delving} to \\ randomly initialize the weights of the model.\end{tabular}                                   \\ \hline
ImageNet                                                                   & \begin{tabular}[c]{@{}l@{}}Model is pretrained on ImageNet dataset\cite{krizhevsky2017imagenet}\\ with millions of annotated natural sense images.\end{tabular}                                                                                  \\ \hline
PneuXray                                                                   & \begin{tabular}[c]{@{}l@{}}A supervised model pretrained using PneuXray\\ dataset\cite{li2019pnet}, which consisits of about 5000\\ chest x-ray images labelled with pneumonia\\ and 5000 x-ray images labelled with normal.\end{tabular} \\ \hline
PLR-L                                                                      & \begin{tabular}[c]{@{}l@{}}A self-supervised pretraining task designed to\\ restore normal CT images from local pixel\\ shuffled\cite{zhou2020models} CT images.\end{tabular}                                              \\ \hline
PLR-G                                                                      & \begin{tabular}[c]{@{}l@{}}A self-supervised pretraining task designed\\ to restore normal CT images from Gaussian \\ blurred ones.\end{tabular}                                                                     \\ \hline
PLR-P                                                                      & \begin{tabular}[c]{@{}l@{}}Our proposed framework, pseudo COVID-19\\ CT images are generated from normal images\\ by pasting lesion-like patches.\end{tabular}                                                     \\ \hline
\end{tabular}
\end{table}

\begin{table}[]
\centering
\caption{The similarity between COVID-19 lesions and pseudo lesions.}
\label{tab:table3}
\begin{tabular}{llll}
\hline
Similarity              & Local shuffing  & Gaussian & Perlin \\ \hline
Cosine                  & 0.0788 & 0.0587 & \textbf{0.0328} \\
JS                      & 0.1363 & 0.1185 & \textbf{0.0923} \\ \hline

\end{tabular}
\end{table}

\begin{figure}[ht]
\centering
\includegraphics[scale=0.45]{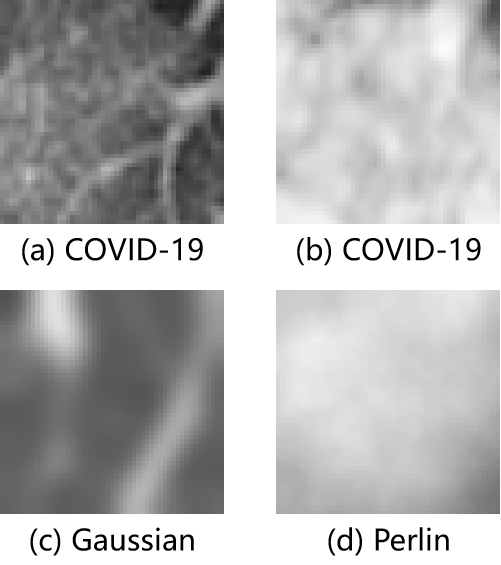}
\caption{Illustration of (a) (b) COVID-19 lesions, (c) pseudo lesion generated by Gaussian blur and (d) pseudo lesion generated by Perlin noise.}
\label{fig7}
\end{figure}

\begin{figure}[ht]
\centering
\includegraphics[scale=0.32]{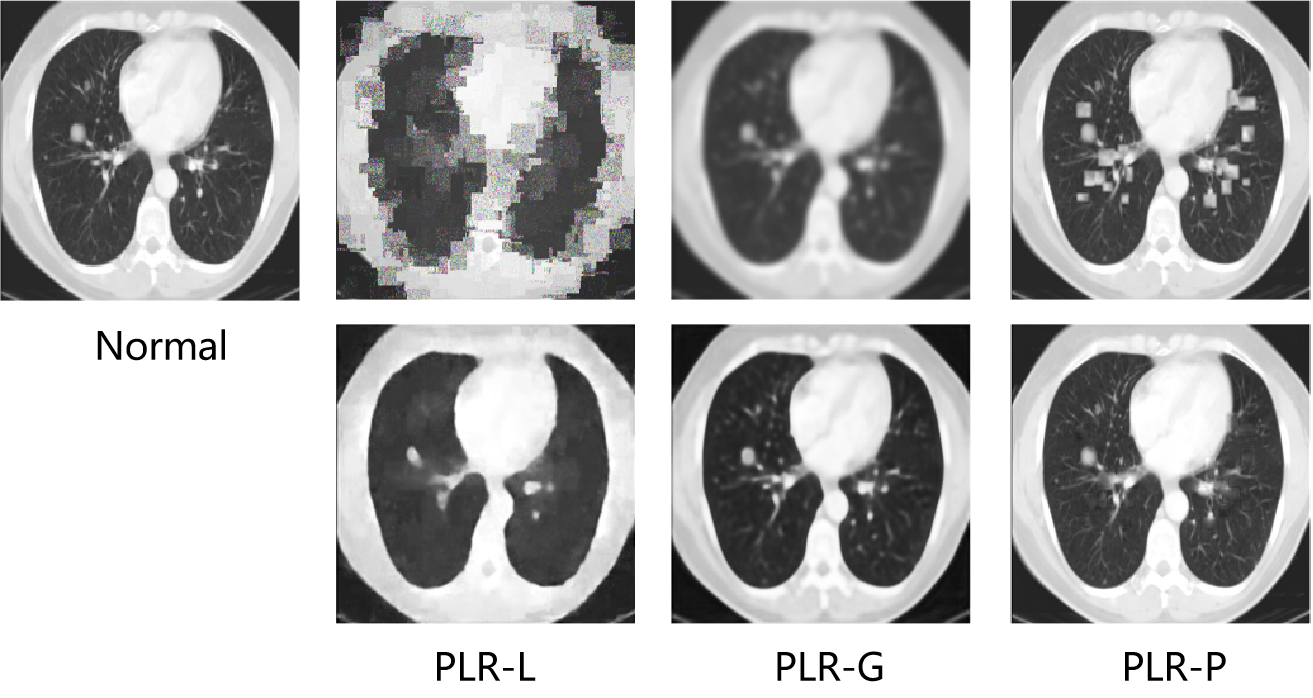}
\caption{Demonstration of the image restoration quality for different self-supervised learning strategies. We illustrate an original normal CT image on the first column. The pairs of pseudo COVID-19 and restored images for PLR-L, PLR-G and PLR-P strategies are listed from second to forth column, respectively.}
\label{fig8}
\end{figure}

\begin{figure}[ht]
\centering
\includegraphics[scale=0.5]{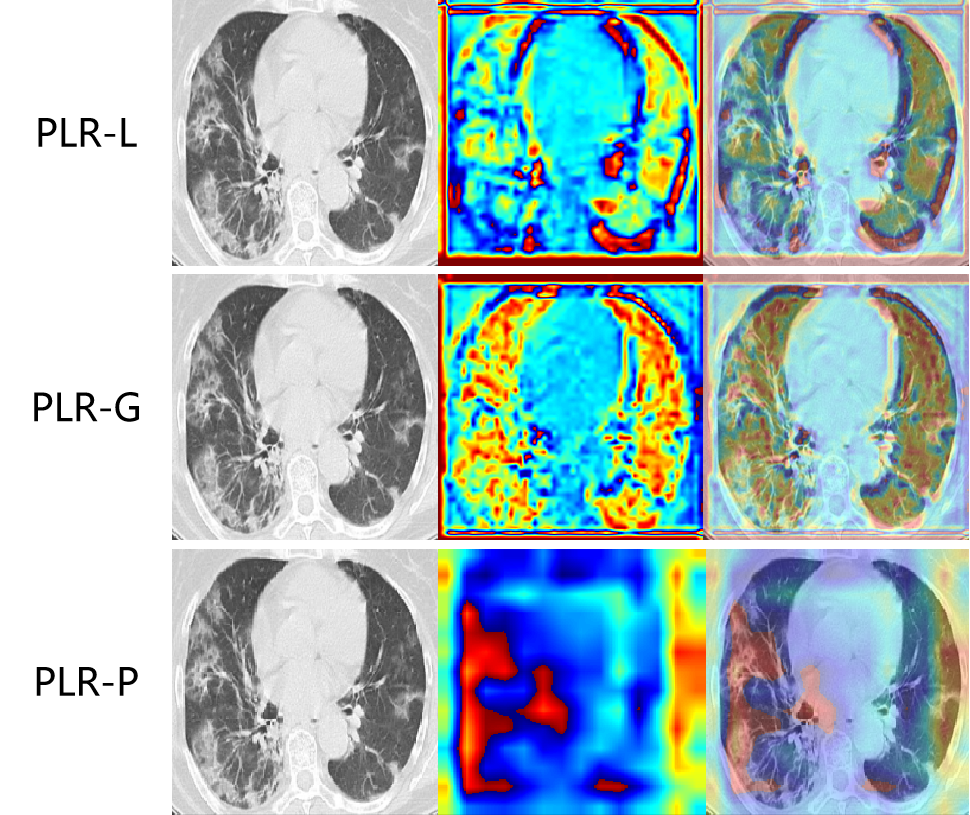}
\caption{Visualization of color maps of conv\_8 layer in different pretrained encoders.}
\label{fig9}
\end{figure}

\begin{figure*}[ht]
\centering
\includegraphics[scale=0.66]{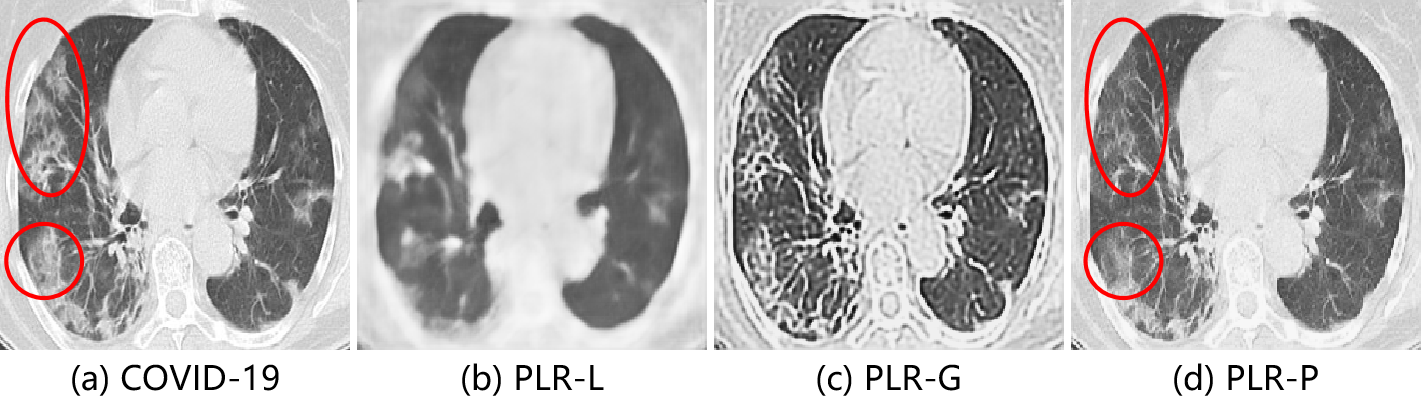}
\caption{The restoration results of different pretrained models for a real COVID-19 CT image.}
\label{figl0}
\end{figure*}

\begin{figure*}[]
\centering
\includegraphics[scale=0.48]{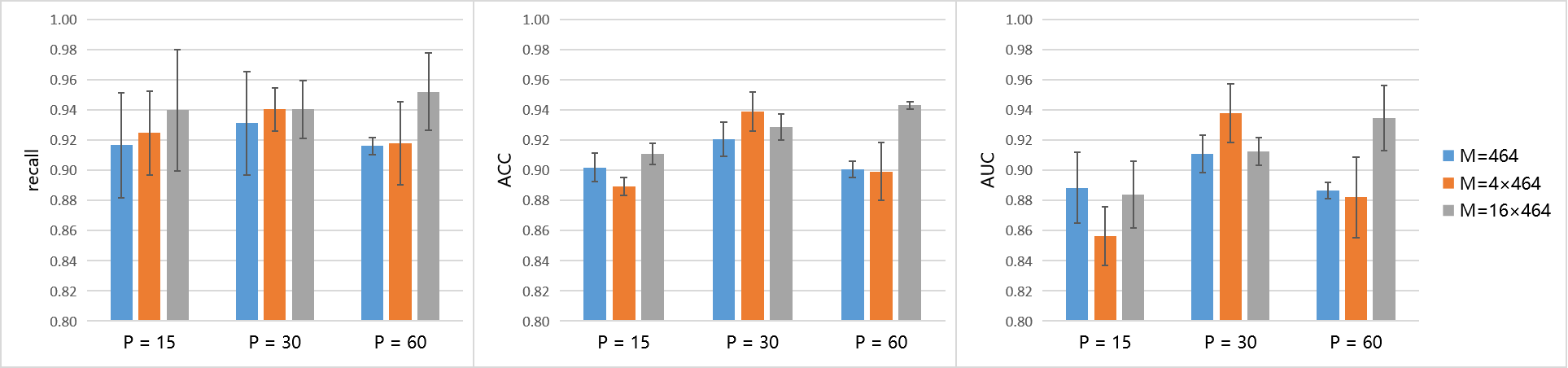}
\caption{Classification performance of PLR-P with different number of PL patches and pseudo COVID-19 images.}
\label{fig11}
\end{figure*}

\subsection{Performance of pseudo-lesion generation and restoration}
In this section, we mainly evaluate the similarity of pseudo lesions with real COVID-19 lesion patches, and the quality of restoration for pseudo COVID-19 images generated by pasting different pseudo lesions.\par

As only image level COVID-19 labels are available in Jinan and SARS-CoV-2 datasets, we collected 1160 COVID-19 lesion patches ($64\times64$) from COVID-19 CT images in UESTC-COVID-19 dataset\cite{wang2020noiserobust}, where lesions of COVID-19 CT images are manually labeled. We also generated 1000 pseudo-lesion patches ($32\times32$) by local pixel shuffling, Gaussian blurring and Perlin noise, as described in section \uppercase\expandafter{\romannumeral3}.B. Each pesudo lesion is matched with all of the COVID-19 lesion patches and the average of all similarity are recorded. The size of all patches are resized to $32\times32$ before similarity measurement. Two common metric, cosine distance and Jensen-Shannon(JS) divergence, are used to measure the similarity between COVID-19 lesions and pseudo lesions. The COVID-19 lesions and pseudo lesions were flattened to vectors in 1024 dimensions to calculate the similarity.\par
Table \uppercase\expandafter{\romannumeral3} lists the average of similarities for different pseudo lesions. As local shuffling randomly shuffle the pixels in the lung area patches, pseudo lesions generated by such a random change appears much different with the COVID-19 lesion pattern, i.e. the similarity of local shuffling with GGO is the lowest among three generators. As shown in Fig. 7, the pseudo lesions generated by both Gaussian and Perlin noises are much more similar with COVID-19 lesion patches, than local shuffling. We can also observed in Table \uppercase\expandafter{\romannumeral3}, the pseudo lesion generated using Perlin noise is more similar than that generated using Gaussian noise. As shown in Fig. 7, the lesions generated by Perlin noise present white and black cloud patterns similar with COVID-19 lesions.\par

\begin{table*}[t]
\centering
\caption{Performance comparison between our method and other pretrained methods on SARS-CoV-2 dataset. We report the Accuracy, Precision, Recall, F1 Score and AUC for each pretrained model.}
\resizebox{0.98\linewidth}{!}{
\begin{tabular}{clcccccccccc}
\hline
\multirow{2}{*}{Fraction of training set} & \multicolumn{1}{c}{\multirow{2}{*}{Methods}} & \multicolumn{2}{c}{Accuracy}      & \multicolumn{2}{c}{Precision}     & \multicolumn{2}{c}{Recall}        & \multicolumn{2}{c}{F1\_score}     & \multicolumn{2}{c}{AUC}           \\ \cline{3-12} 
                                          & \multicolumn{1}{c}{}                         & Highest         & Average         & Highest         & Average         & Highest         & Average         & Highest         & Average         & Highest         & Average         \\ \hline
\multirow{6}{*}{100\%}                    & RI                                           & 0.8876          & 0.8708          & 0.9529          & 0.9453          & 0.9144          & 0.8767          & 0.9233          & 0.9093          & 0.8684          & 0.8654          \\
                                          & ImageNet                                     & 0.8911          & 0.8788          & 0.9594          & 0.9497          & 0.8934          & 0.8831          & 0.9240          & 0.9152          & 0.8891          & 0.8748          \\
                                          & PneuXray                                     & 0.9031          & 0.8927          & 0.9786          & 0.9721          & 0.8982          & 0.8804          & 0.9314          & 0.9239          & 0.9166          & 0.9041          \\
                                          & PLR-L                                        & 0.8888          & 0.8876          & 0.9647          & 0.9476          & 0.9273          & 0.8988          & 0.9228          & 0.9221          & 0.8950          & 0.8772          \\
                                          & PLR-G                                        & 0.8983          & 0.8935          & 0.9701          & 0.9442          & 0.9402          & 0.9111          & 0.9284          & 0.9269          & 0.9059          & 0.8772          \\
                                          & PLR-P                                        & \textbf{0.9533} & \textbf{0.9390} & \textbf{0.9899} & \textbf{0.9767} & \textbf{0.9499} & \textbf{0.9402} & \textbf{0.9678} & \textbf{0.9580} & \textbf{0.9595} & \textbf{0.9379} \\ \hline
\multirow{6}{*}{50\%}                     & RI                                           & 0.8696          & 0.8684          & 0.9568          & 0.9553          & 0.8659          & 0.8627          & 0.9077          & 0.9066          & 0.8744          & 0.8737          \\
                                          & ImageNet                                     & 0.8840          & 0.8780          & 0.9638          & 0.9539          & 0.8966          & 0.8781          & 0.9196          & 0.9141          & 0.8836          & 0.8780          \\
                                          & PneuXray                                     & 0.9079          & 0.9037          & 0.9428          & 0.9425          & 0.9321          & 0.9265          & 0.9374          & 0.9344          & 0.8854          & 0.8826          \\
                                          & PLR-L                                        & 0.8900          & 0.8870          & 0.9444          & 0.9434          & 0.9047          & 0.9015          & 0.9241          & 0.9220          & 0.8763          & 0.8736          \\
                                          & PLR-G                                        & 0.8971          & 0.8858          & 0.9405          & 0.9262          & 0.9192          & 0.9192          & 0.9297          & 0.9226          & 0.8767          & 0.8548          \\
                                          & PLR-P                                        & \textbf{0.9294} & \textbf{0.9282} & \textbf{0.9729} & \textbf{0.9690} & \textbf{0.9386} & \textbf{0.9330} & \textbf{0.9517} & \textbf{0.9506} & \textbf{0.9268} & \textbf{0.9239} \\ \hline
\multirow{6}{*}{30\%}                     & RI                                           & 0.8170          & 0.8140          & 0.8385          & 0.8297          & 0.9628          & 0.9427          & 0.8862          & 0.8824          & 0.7078          & 0.6949          \\
                                          & ImageNet                                     & 0.7895          & 0.7877          & 0.8235          & 0.8158          & 0.9386          & 0.9217          & 0.8685          & 0.8654          & 0.6758          & 0.6636          \\
                                          & PneuXray                                     & 0.8541          & 0.8487          & 0.9041          & 0.9001          & 0.8982          & 0.8950          & 0.9011          & 0.8975          & 0.8132          & 0.8058          \\
                                          & PLR-L                                        & 0.8194          & 0.8152          & 0.8385          & 0.8253          & \textbf{0.9838} & \textbf{0.9532} & 0.8897          & 0.8841          & 0.7078          & 0.6874          \\
                                          & PLR-G                                        & \textbf{0.8612} & \textbf{0.8583} & 0.9116          & \textbf{0.8980} & 0.9257          & 0.9128          & \textbf{0.9057} & \textbf{0.9051} & 0.8255          & \textbf{0.8078} \\
                                          & PLR-P                                        & 0.8505          & 0.8499          & \textbf{0.9214} & 0.8908          & 0.9532          & 0.9120          & 0.9042          & 0.8998          & \textbf{0.8294} & 0.7924          \\ \hline
\multirow{6}{*}{10\%}                     & RI                                           & 0.8230          & 0.8176          & 0.8213          & 0.8206          & 0.9725          & 0.9645          & 0.8905          & 0.8867          & 0.6844          & 0.6816          \\
                                          & ImageNet                                     & 0.8194          & 0.8188          & 0.8382          & 0.8343          & 0.9483          & 0.9427          & 0.8854          & 0.8851          & 0.7104          & 0.7041          \\
                                          & PneuXray                                     & 0.7943          & 0.7931          & 0.8108          & 0.8048          & 0.9612          & 0.9515          & 0.8724          & 0.8719          & 0.6576          & 0.6463          \\
                                          & PLR-L                                        & 0.8038          & 0.8026          & 0.8090          & 0.8060          & 0.9742          & 0.9661          & 0.8803          & 0.8788          & 0.6564          & 0.6513          \\
                                          & PLR-G                                        & 0.8170          & 0.8152          & 0.8284          & 0.8229          & 0.9693          & 0.9564          & 0.8869          & 0.8846          & 0.6929          & 0.6844          \\
                                          & PLR-P                                        & \textbf{0.8539} & \textbf{0.8517} & \textbf{0.8436} & \textbf{0.8386} & \textbf{0.9935} & \textbf{0.9903} & \textbf{0.9125} & \textbf{0.9082} & \textbf{0.7341} & \textbf{0.7233} \\ \hline
\end{tabular}}
\end{table*}

Now we visually compare the quality of restoration for pseudo COVID-19 images generated by different pseudo lesions. Fig. 8 shows an example normal images, the pseudo COVID-19 images generated by three different pseudo lesions and the restored normal CT images. While PLR-L restored the main outline of lung, the texture information is missing. PLR-G restored the normal CT image from pseudo COVID-19 image by enhancing the details of blurred image. In the last column, we can observe that PLR-P restored the pseudo lesions to normal lung context patches successfully.\par
Fig. 9 shows the visualization of conv\_8 layers for the restoration network trained using COVID-19 images generated by three different pseudo lesions. It can be observed that PLR-L mainly focuses on the edge of lung and PLR-G focuses on the vessels distributing in the lung region. It can also be observed that PLR-P focuses on the COVID-19 lesions, which suggests that model pretrained by PLR-P presenting promising sensitivity to COVID-19 lesions.\par
Furthermore, we directly apply the restoration networks to real COVID-19 images to see the results. Fig. 10 shows an example COVID-19 image and the output of the three restoration networks trained by pseudo COVID-19 images generated using different pseudo lesions. As illustrated in Fig. 10., While the image restored by PLR-L is very vague and looks very different with the original image, the images restored by PLR-G and PLR-P look much more clear. However, the model of PLR-G tends to focus on the contrast of local structures like vessels and the COVID-19 lesion pattern in the red circle region is not fully restored, though contrast of whole image is enhanced. The image restored using PLR-P is much more similar with the original image and most of the COVID-19 lesion patterns have been removed. 

\begin{table*}[]
\centering
\caption{Performance comparison between different methods on Jinan COVID-19 Dataset (Note, / means that the results are not reported by that method.).}
\resizebox{0.98\linewidth}{!}{
\begin{tabular}{clcccccccccc}
\hline
\multirow{2}{*}{Fraction of training set} & \multicolumn{1}{c}{\multirow{2}{*}{Methods}} & \multicolumn{2}{c}{Accuracy}      & \multicolumn{2}{c}{Precision}     & \multicolumn{2}{c}{Recall}        & \multicolumn{2}{c}{F1\_score}     & \multicolumn{2}{c}{AUC}           \\ \cline{3-12}
                                          & \multicolumn{1}{c}{}                         & Highest         & Average         & Highest         & Average         & Highest         & Average         & Highest         & Average         & Highest         & Average         \\ \hline
\multirow{7}{*}{100\%}                    & \multicolumn{1}{c}{DANet(Hu et al.)}         & 0.9269          & 0.8697          & /               & /               & 0.9800          & 0.8980          & /               & /               & 0.9796          & 0.9293          \\
                                          & RI                                           & 0.9495          & 0.9394          & 0.9412          & 0.9235          & 0.9600          & 0.9600          & 0.9505          & 0.9413          & 0.9494          & 0.9392          \\
                                          & ImageNet                                     & 0.9495          & 0.9192          & 0.9245          & 0.9003          & \textbf{0.9800} & 0.9467          & 0.9515          & 0.9226          & 0.9492          & 0.9189          \\
                                          & PneuXray                                     & 0.9293          & 0.9259          & 0.9057          & 0.9158          & 0.9600          & 0.9400          & 0.9320          & 0.9276          & 0.9290          & 0.9258          \\
                                          & PLR-L                                        & 0.9596          & 0.9360          & 0.9792          & 0.9418          & 0.9400          & 0.9333          & 0.9592          & 0.9367          & 0.9598          & 0.9361          \\
                                          & PLR-G                                        & 0.9697          & 0.9327          & 0.9796          & 0.9291          & 0.9600          & 0.9400          & 0.9697          & 0.9337          & 0.9698          & 0.9326          \\
                                          & PLR-P                                        & \textbf{0.9798} & \textbf{0.9663} & \textbf{0.9800} & \textbf{0.9682} & \textbf{0.9800} & \textbf{0.9667} & \textbf{0.9800} & \textbf{0.9669} & \textbf{0.9798} & \textbf{0.9663} \\ \hline
\multirow{6}{*}{50\%}                     & RI                                           & 0.7576          & 0.7576          & 0.7955          & 0.7955          & 0.7000          & 0.7000          & 0.7447          & 0.7447          & 0.7582          & 0.7582          \\
                                          & ImageNet                                     & 0.7374          & 0.7374          & 0.7857          & 0.7857          & 0.6600          & 0.6600          & 0.7174          & 0.7174          & 0.7382          & 0.7382          \\
                                          & PneuXray                                     & 0.7778          & 0.7576          & 0.8043          & 0.7950          & 0.7400          & 0.7000          & 0.7708          & 0.7441          & 0.7782          & 0.7582          \\
                                          & PLR-L                                        & 0.7475          & 0.7475          & 0.8049          & 0.8049          & 0.6600          & 0.6600          & 0.7253          & 0.7253          & 0.7484          & 0.7484          \\
                                          & PLR-G                                        & \textbf{0.8081} & 0.7576          & 0.8298          & \textbf{0.8149} & \textbf{0.7800} & 0.6700          & \textbf{0.8041} & 0.7315          & \textbf{0.8084} & 0.7585          \\
                                          & PLR-P                                        & 0.7879          & \textbf{0.7778} & 0.8085          & 0.7866          & 0.7600          & \textbf{0.7700} & 0.7835          & \textbf{0.7779} & 0.7882          & \textbf{0.7779} \\ \hline
\multirow{6}{*}{30\%}                     & RI                                           & 0.7273          & 0.7273          & 0.7255          & 0.7255          & 0.7400          & 0.7400          & 0.7327          & 0.7327          & 0.7271          & 0.7271          \\
                                          & ImageNet                                     & 0.6970          & 0.6970          & 0.6613          & 0.6613          & 0.8200          & 0.8200          & 0.7321          & 0.7321          & 0.6957          & 0.6957          \\
                                          & PneuXray                                     & 0.7475          & 0.7273          & \textbf{0.7907} & 0.7444          & 0.6800          & 0.7100          & 0.7312          & 0.7248          & 0.7482          & 0.7275          \\
                                          & PLR-L                                        & 0.7576          & 0.7374          & 0.7097          & 0.6997          & \textbf{0.8800} & \textbf{0.8400} & \textbf{0.7857} & 0.7632          & 0.7563          & 0.7363          \\
                                          & PLR-G                                        & 0.7273          & 0.7273          & 0.7170          & 0.7170          & 0.7600          & 0.7600          & 0.7379          & 0.7379          & 0.7269          & 0.7269          \\
                                          & PLR-P                                        & \textbf{0.7677} & \textbf{0.7576} & 0.7547          & 0.7410          & 0.8000          & 0.8000          & 0.7767          & \textbf{0.7693} & \textbf{0.7673} & \textbf{0.7571} \\ \hline
\end{tabular}}
\end{table*}

\subsection{Selection of PLR Hyper-parameters}
In this section, we investigated two hyper-parameters in our PLR framework, the number of PL patches $(P)$ pasted to each CT images and number of pseudo COVID-19 images $(M)$ employed in pretraining, based on the validation set of SARS-CoV-2 dataset. Fig. 11 shows evolution of the classification performance on the validation set when $P \in \{ p,p\times2,p\times4\mid p=15 \}$ and $M \in \{ n,n\times4,n\times16\mid n=464 \}$. As there are 464 normal CT images in the training set of SARS-CoV-2 dataset, we generate $N$ (multiply of 464) pairs of normal and pseudo COVID-19 images to train the encoder and test its performance using the validation set. As shown in Fig.11, the performance of PLR-P generally improve when $P$ increases from 15 to 30, for various $M$. However, the performance of PLR-P with $M=4\times464$ and $16\times464$ starts to decrease when $P$ increases from 30 to 60. Considering both efficiency and performance, we choose $P=30$ and $M=1856$ in the following experiments.

\subsection{Diagnosis Performance on SARS-CoV-2 dataset}
We then study the effectiveness of our proposed PLR framework for COVID-19 diagnosis. To compare different pretraining methods in fairness, we repeat three experiments independently for each method. Table \uppercase\expandafter{\romannumeral4} summarizes the results of different pretraining strategies on SARS-CoV-2 datasets. As shown in Table \uppercase\expandafter{\romannumeral4}, PLR-P outperformed other methods significantly in all five metrics when different ratio of the training set was used for finetuning. Specifically, PLR-P achieved the highest accuracy of 0.9533, 0.9294, 0.8539 and AUC of 0.9595, 0.9268 and 0.7341 among all of the pretraining strategies when using 100\%, 50\% and 10\% of the training set for finetuning, respectively. From Table \uppercase\expandafter{\romannumeral4} we can also observe that our method significantly improved model’s performance, compared to the random initialization method. The PLR-P pretrained network achieves average recall and AUC of 0.9402 and 0.9379, which are 6.35\% and 7.25\% higher than that of random initialized network, respectively. Similar improvements are also observed for accuracy, precision and F1 score. It seems that PLR-P model learns more efficient features from lesion-like patterns for COVID-19 diagnosis.\par
We notice that the performance improvement is highly related to the diversity of the pretraining datasets and strategies. Due to the gap between natural images and medical data, the improvement i.e., +0.8\% and +0.94\%, of average accuracy and AUC yielded by ImageNet pretraining is limited. PneuXray model, pretrained on 10,000 x-ray images for pneumonia diagnosis improves average accuracy and AUC by 2.19\% and 3.87\%. Similarly, PLR-P pretrained on lesion-like patches benefit the COVID-19 diagnosis, i.e. PLR-P achieved higher AUC, recall, accuracy, etc. than other self-supervised models in general.\par
To better demonstrated the superiority of the proposed method when annotated training data is limited, an experiment is conducted to evaluate the classification performance of PLR-P model with different amounts of labeled data used for finetuning (i.e., 10\%, 30\%, 50\%). It can be observed from Table \uppercase\expandafter{\romannumeral4} that our PLR-P can effectively deal with the situation when few labeled training samples are available. PLR-P trained using 50\% labeled data, achieves an accuracy of 0.9294, precision of 0.9729, recall of 0.9386, F1 score of 0.9517 and AUC of 0.9268, which is even superior to other methods trained with 100\% training data.\par
We also notice that, when 30\% annoated data is used for finetuning, PLR-G model achieves 1.07\% and 0.15\% higher performance than PLR-P model on highest accuracy and F1 score, respectively. The underlying reason may be that, PLR-G learnt high-level sematic features, e.g. lung tissue and vessel patterns, which may benefit COVID-19 diagnosis.\par
Moreover, when networks are finetuned using only 10\% of lablelled data, PLR-P model achieves a relatively high recall of 0.9935, with precision of 0.8436 and AUC of 0.7341. As COVID-19 is a highly contagious disease, isolating COVID-19 patients as early as possible is one of the keys to vanquish the pandemic.\par

\subsection{Diagnosis Performance on Jinan COVID-19 dataset}
To further evaluate the effectiveness of the proposed method, we also employed Jinan COVID-19 dataset for evaluation. For this dataset, we used the same experimental settings as those in the SARS-CoV-2 dataset. Table \uppercase\expandafter{\romannumeral5} shows the highest and average performance of networks initialized with different pretraining methods when 100\%, 50\% and 30\% of the annotated training set was used for finetuning. One could observe from the Table \uppercase\expandafter{\romannumeral5} that, the proposed PLR-P pretraining method achieved the best performance on all five evaluation metrics when whole training set is used for finetuning. The PLR-P model pretrained with 303 unlabeled images achieves above 4\% higher accuracy, F1 score and AUC than the model pretrained on ImageNet with millions of labelled images. The results demonstrate that the deep features learned from lesion-like patches benefit COVID-19 CT images screening.\par
It could also be observed that models pretrained by self-supervised learning tasks outperformed other pretrained models when using whole training set for finetuning. Specifically, models pretrained on ImageNet and PneuXray only achieved the highest AUC of 0.9492 and 0.929, while the self-supervised strategies achieved the highest AUC in range of [0.9598, 0.9798]. Similar trends of improvement are observed for accuracy, precision and F1 score. The results suggest that pretraining with appropriate strategies can help deep model learn features related to downstream tasks. Models pretrained by PLR tasks with using a small amount of unlabelled images extract more robust features than models pretrained by a mass of labelled images, e.g. ImageNet and PneuXray.\par
As the size of training set used for finetuning decreases, the performances of all the models declined. In this case, PLR methods still achieves the best Accuracy, F1 score and AUC. When the perecentage of training set decrease from 50\% to 30\%, average AUC of model pretrained on ImageNet and PneuXray decreased 4.25\% and 3.07\%, respectively, while that of proposed PLR-P model only decreased 2.08\%.\par
Furthermore, we compared three self-supervised pretrained model, i.e. PLR-L, PLR-G and PLR-P. As shown in Table \uppercase\expandafter{\romannumeral3}, the pseudo lesions employed in PLR-L demonstrated the lowest similarity with real lesion, which lead to the worst performance.\par

\section{Conclusion}
This paper proposes a novel self-supervised learning framework, PLR, for COVID-19 diagnosis. Our method pretrains the model using an image restoration task designed specially to learn COVID-19 lesion related features for COVID-19 diagnosis. We evaluated our method on two publicly available large-scale datasets by comparing with various pretraining strategies. Experimental results demonstrate the superiority of our self-supervised method. Especially, the proposed method achieves excellently high recall for COVID-19 patients with resonable precision.\par

\end{document}